\documentclass{PoS_arXiv}

\usepackage{amsmath,amssymb,amsfonts,color,graphicx,cite,color,soul}

\title{A Higgs Boson at 96 GeV?!}

\ShortTitle{A Higgs Boson at 96 GeV?!}

\author{\speaker{S.~Heinemeyer}\\ 
        Instituto de F\'isica Te\'orica, (UAM/CSIC), Universidad
  Aut\'onoma de Madrid,\\  
Cantoblanco, E-28049 Madrid, Spain\\
Campus of International Excellence UAM+CSIC, Cantoblanco, E-28049,
Madrid, Spain\\
Instituto de F\'isica de Cantabria (CSIC-UC), E-39005 Santander, Spain\\
        E-mail: \email{Sven.Heinemeyer@cern.ch}}

\author{T.~Stefaniak\\
        DESY, Notkestra{\ss}e 85, D-22607 Hamburg, Germany\\
        E-mail: \email{tim.stefaniak@desy.de}}

\abstract{We briefly summarize some searches for Higgs bosons with a mass of 
$\mphi \lsim 110 \gev$ at LEP and the LHC. We discuss a possible
signal in the diphoton decay mode at $\mphi \sim 96 \gev$ as reported by
CMS, together with a $\sim 2\,\sig$ hint in the $b \bar b$ final state
at LEP. We briefly review possible interpretation of such a new particle
in various BSM models. We focus on possible explanations as reported
within the NMSSM and the \mnSSM. Conclusions for future collider
projects are briefly outlined.
}

\FullConference{
Prospects for Charged Higgs Discovery at Colliders - CHARGED2018\\
		25-28 September 2018\\
		Uppsala, Sweden}

\newcommand{\lsim}
{\;\raisebox{-.3em}{$\stackrel{\displaystyle <}{\sim}$}\;}

\newcommand\tb{\tan\beta}

\newcommand\ReDiag{\mathop{%
  \raise .5pt\hbox{[}%
  \widetilde{\mathrm{Re}}%
  \raise .5pt\hbox{]}}}
\newcommand\ReOffDiag{\mathop{%
  \raise .5pt\hbox{$\llbracket$}%
  \widetilde{\mathrm{Re}}%
  \raise .5pt\hbox{$\rrbracket$}}}

\newcommand\MHp{M_{H^\pm}}

\newcommand\Ab{A_b}
\newcommand\At{A_t}

\newcommand\refeq[1]{Eq.~(\ref{#1})}

\newcommand\refta[1]{Tab.~\ref{#1}}
\newcommand\refse[1]{Sect.~\ref{#1}}

\newcommand\citere[1]{Ref.~\cite{#1}}
\newcommand\citeres[1]{Refs.~\cite{#1}}

\newcommand{\CP}{{\cal CP}}
\newcommand{\cp}{{\CP}}

\newcommand{\tev}{\,\, \mathrm{TeV}}
\newcommand{\gev}{\,\, \mathrm{GeV}}

\newcommand{\br}{\text{BR}}

\newcommand{\sig}{\sigma}

\def\reffi#1{\mbox{Fig.~\ref{#1}}}

\def\Ga{\Gamma}
\def\ga{\gamma}

\def\la{\lambda}

\definecolor{Lightblue}{cmyk}{0.9,0.1,0.1,0.3}
\definecolor{dgelborange}{cmyk}{0.,0.3,0.5, 0.}
\definecolor{Lila}{rgb}{0.5,0.,1}

\graphicspath{{figs/}}

\newcommand{\mphi}{m_\phi}
\newcommand{\mnSSM}{\ensuremath{\mu\nu\mathrm{SSM}}}

\begin{document}


\section{Introduction}
\label{sec:intro}

In the year 2012 the ATLAS and CMS collaborations have discovered a new
particle that -- within theoretical and experimental uncertainties -- is
consistent with the existence of a Standard-Model~(SM) Higgs boson at a mass
of~$\sim 125 \gev$~\cite{Aad:2012tfa,Chatrchyan:2012xdj,Khachatryan:2016vau}.
No conclusive signs of physics beyond the~SM have been found so far at the LHC.
However, the measurements of Higgs-boson couplings, which are known
experimentally to a precision of roughly $\sim 20\%$, leave room for
Beyond Standard Model (BSM) interpretations. Many BSM models possess
extended Higgs-boson sectors. Consequently, 
one of the main tasks of the LHC Run~II and beyond will be to determine whether
the observed scalar boson forms part of the Higgs sector of an extended
model.

Motivated by the Hierarchy Problem, Supersymmetry~(SUSY)
extensions of the~SM play a prominent role in the exploration of new
physics. 
SUSY doubles the particle degrees of freedom by predicting
two scalar partners for all SM fermions, as well as fermionic partners
to all SM bosons. The simplest SUSY extension is the Minimal Supersymmetric
Standard Model (MSSM)~\cite{Nilles:1983ge,Haber:1984rc}.
In contrast to the single Higgs doublet of the SM, in the MSSM 
two Higgs doublets, $H_u$ and $H_d$, are required. 
In the $\cp$ conserving case the MSSM Higgs sector 
consists of two $\cp$-even, one
$\cp$-odd and two charged Higgs bosons. The light (or the heavy)
$\cp$-even MSSM Higgs boson can be interpreted as the signal discovered
at $\sim 125 \gev$~\cite{Heinemeyer:2011aa} (see
\citeres{Bechtle:2016kui,Bahl:2018zmf} for recent updates). 
   
Going beyond the MSSM, a well-motivated extension is given by
the Next-to-MSSM (NMSSM), see
e.g.~\cite{Ellwanger:2009dp,Maniatis:2009re} for reviews. 
The NMSSM provides a solution for the so-called ``$\mu$ problem'' by
naturally associating an adequate scale to the $\mu$ parameter appearing
in the MSSM superpotential~\cite{Ellis:1988er,Miller:2003ay}.
In the NMSSM a new singlet superfield is introduced, which only couples to the
Higgs- and sfermion-sectors, giving rise to an effective $\mu$-term,
proportional to the vacuum expectation value (vev) of the scalar singlet.
In the $\cp$ conserving case the NMSSM Higgs sector consists of
three $\cp$-even Higgs bosons,
$h_i$ ($i = 1,2,3$), two $\cp$-odd Higgs bosons, $a_j$ ($j = 1,2$),
and the charged Higgs boson pair $H^\pm$. 
In the NMSSM the
lightest but also the second lightest $\cp$-even Higgs boson
can be interpreted as the signal observed at about $125~\gev$, see,
e.g., \cite{King:2012is,Domingo:2015eea}.

A natural extension of the NMSSM is the \mnSSM, in which the 
singlet superfield is interpreted as a right-handed neutrino 
superfield~\cite{LopezFogliani:2005yw,Escudero:2008jg} (here we focus on
the ``one generation case''),
see \citeres{Munoz:2009an,Munoz:2016vaa,Ghosh:2017yeh} for reviews. 
The \mnSSM\ is the simplest extension of the MSSM that can provide 
massive neutrinos through a see-saw mechanism at the electroweak scale.
A Yukawa coupling for the right-handed neutrino of the order of the
electron Yukawa coupling is introduced that induces the explicit
breaking of $R$-parity. One consequences is that there is no
lightest stable SUSY particle anymore. Nevertheless, the model can still
provide a dark matter candidate with a gravitino that has a life time longer  
than the age of the observable 
universe~\cite{Choi:2009ng,GomezVargas:2011ph,Albert:2014hwa,Gomez-Vargas:2016ocf}. The explicit violation of lepton number and lepton flavor can modify the 
spectrum of the neutral and charged fermions in comparison to the NMSSM.
The three families of charged leptons will mix with the chargino and 
the Higgsino and form five massive charged fermions. Within the scalar sector,
due to $R$-parity breaking, the 
left- and right-handed sneutrinos  
will mix with the doublet Higgses and form six massive $\cp$-even 
and five massive $\cp$-odd states, assuming that there is no 
$\cp$-violation. Also in the \mnSSM\ the signal at $\sim 125 \gev$ can
be interpreted as the lightest or the second lightest $\cp$-even scalar.
As SUSY models, but also other BSM Higgs-boson sector extensions can
possess a scalar below $125 \gev$ the search for such light scalars is
of high importance at the LHC.


\section{Experimental data}
\label{sec:exp}

Searches for Higgs bosons below $125 \gev$ have been performed at LEP,
the Tevatron and the LHC. 
LEP reported a $2.3\,\sigma$ local excess
observed in the~$e^+e^-\to Z(H\to b\bar{b})$
searches\,\cite{Barate:2003sz}, which would be consistent with a
scalar mass of~$\sim 98 \gev$ (but due to the final state the 
mass resolution is rather coarse). The ``excess'' corresponds to 
\begin{equation}
\mu_{\rm LEP}=\frac{\sigma\left( e^+e^- \to Z \phi \to Zb\bar{b} \right)}
			   {\sigma^{SM}\left( e^+e^- \to Z H_{\rm SM}
			   		\to Zb\bar{b} \right)}
			  = 0.117 \pm 0.057 \; ,
\label{muLEP}
\end{equation}
where the signal strength $\mu_{\rm LEP}$ is the
measured cross section normalized to the SM expectation,
with the SM Higgs-boson mass at $\sim 96\gev$.
The value for $\mu_{\rm LEP}$ was extracted in \citere{Cao:2016uwt}
using methods described in \citere{Azatov:2012bz}.

\begin{figure}[htb!]
  \centering
  \includegraphics[width=0.6\textwidth]{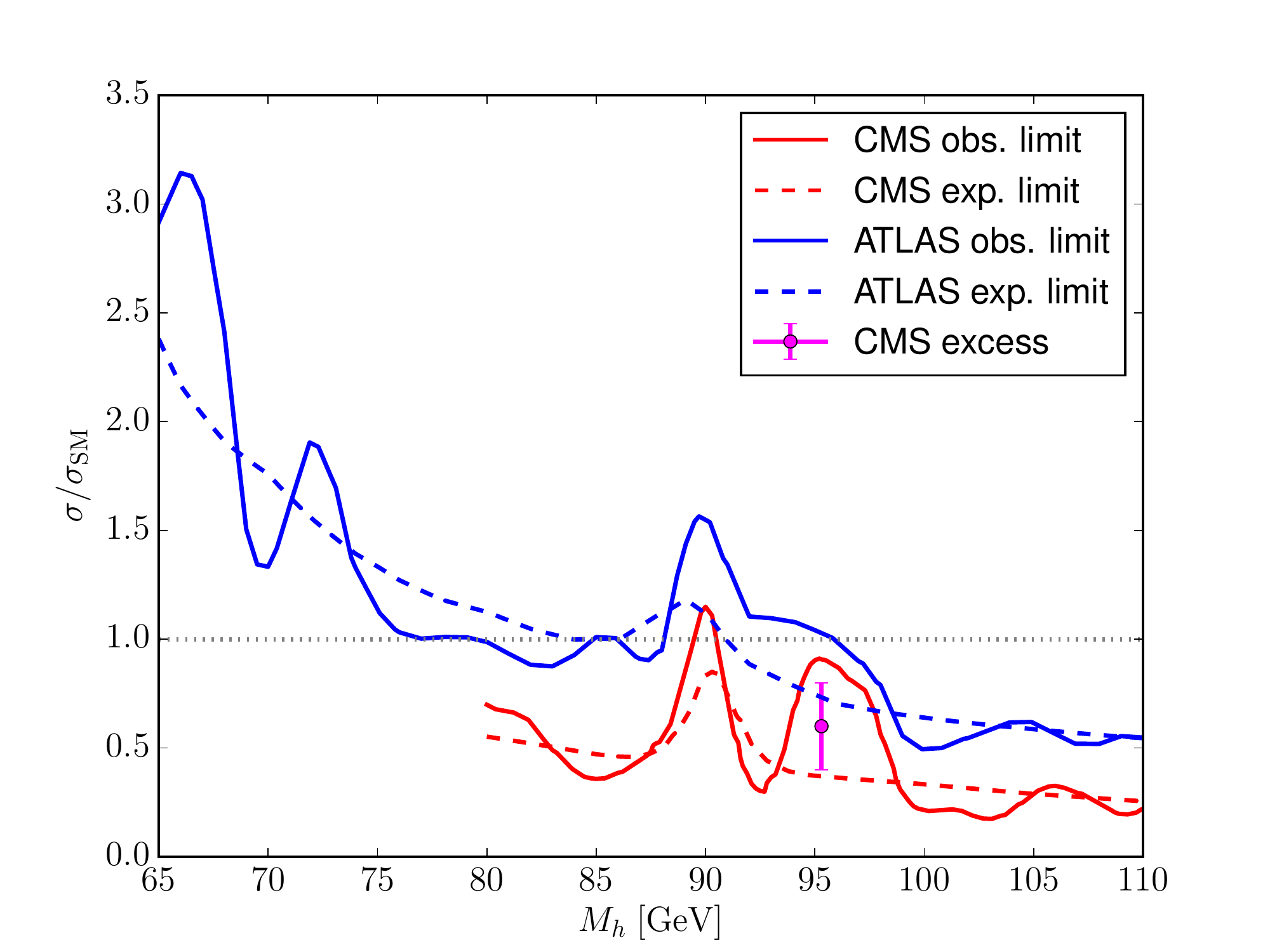}
\vspace{-1em}
  \caption{Limits on the cross section $gg \to \phi \to \ga\ga$
    normalized to the SM value as a function of $\mphi (\equiv M_h)$.
    Compared are the expected (dashed) and observed
    (solid) limits from CMS (red) and ATLAS (blue). Shown in magenta is
  $\mu_{\rm CMS} = 0.6 \pm 0.2$.}
\label{fig:CMS-ATLAS}
\end{figure}

Interestingly, recent CMS~Run\,II
results\,\cite{CMS:2017yta} for Higgs searches in the diphoton
final state show a local excess of~$\sim 3\,\sigma$ around
$\sim 96 \gev$, with a similar excess
of~$2\,\sigma$ in the Run\,I data at a comparable mass.  
In this case the ``excess'' corresponds to (combining 7, 8 and $13 \tev$ data)
\begin{equation}
\mu_{\rm CMS}=\frac{\sigma\left( gg \to \phi \to \gamma\gamma \right)}
         {\sigma^{\rm SM}\left( gg \to H_{\rm SM} \to \gamma\gamma \right)}
     = 0.6 \pm 0.2 \; .
\label{muCMS}
\end{equation}
First Run\,II~results from~ATLAS
with~$80$\,fb$^{-1}$ in the~$\ga\ga$~searches below~$125$\,GeV were
recently published~\cite{ATLAS:2018xad}. No significant excess above
the~SM~expectation was observed in the mass range between $65$~and
$110 \gev$. 
However, the limit on cross section times branching ratio obtained in the
diphoton final state by ATLAS is not only well above $\mu_{\rm CMS}$,
but even weaker than the corresponding upper limit obtained by CMS at 
$\sim 96 \gev$. This is illustrated in \reffi{fig:CMS-ATLAS}, where we
compare the expected (dashed) and observed (solid) limits in the
$gg \to \phi \to \ga\ga$ channel (normalized to the SM value)
as reported by CMS (red) and ATLAS
(blue) as a function of $\mphi$. 
Shown in magenta is $\mu_{\rm CMS}$ of \refeq{muCMS}. The ``weaker''
expected (and observed) exclusion around $91 \gev$ corresponds to the
$Z$~peak, where a larger background is expected.

\section{BSM models}
\label{sec:bsm}

Several analyses attempted to explain the {\em combined} ``excess'' of
LEP and CMS in a variety of BSM models%
\footnote{More analyses attempted to explain one of the two
  ``excesses'', but we will not discuss these further here.}%
. To our knowledge explanations exist in the following frameworks
(see also \cite{Heinemeyer:2018jcd}):
\begin{itemize}
\item Higgs singlet with additional vector-like
  matter, as well as Type-I 2HDM~\cite{Fox:2017uwr}.
\item Radion model~\cite{Richard:2017kot}.
\item Type-I 2HDM with a moderately-to-strongly
  fermiophobic $\cp$-even Higgs~\cite{Haisch:2017gql}.
\item \mnSSM~\cite{Biekotter:2017xmf}, as will be discussed in
  \refse{sec:mnSSM}.
\item Higgs associated with the breakdown of an $U(1)_{L_\mu L_\tau}$
  symmetry~\cite{Liu:2018xsw}. 
\item NMSSM~\cite{Domingo:2018uim}, as will be discussed in 
  \refse{sec:NMSSM}.
\item Higgs inflation inspired $\mu$NMSSM~\cite{Hollik:2018yek}.
\end{itemize}
On the other hand, in the MSSM the CMS excess cannot be
realized~\cite{Bechtle:2016kui}.


\subsection{The NMSSM solution}
\label{sec:NMSSM}

The results in this section are based on \citere{Domingo:2018uim}.
Within the NMSSM a natural candidate to explain the LEP ``excess'' 
consists in a mostly singlet-like Higgs with a doublet
component of about~$10\%$~(mixing squared).
Relatively large Higgs branching fractions
into~$\ga\ga$ are possible due to the three-state mixing, in
particular when the effective Higgs coupling to~$b\bar{b}$ becomes
small, see e.g.~\citeres{Ellwanger:2010nf,Benbrik:2012rm}. 
In our numerical analysis we display the
quantities~$\xi_b$ and~$\xi_{\ga}$, defined as follows:
\begin{align}
& \xi_b\equiv\frac{\Ga(h_1\to ZZ)\cdot \br(h_1\to b\bar{b})}{\Ga(H_{\mbox{\tiny SM}}(M_{h_1})\to ZZ)\cdot \br(H_{\mbox{\tiny SM}}(M_{h_1})\to b\bar{b})}\sim\frac{\sig(e^+e^-\to Z (h_1\to b\bar{b}))}{\sig(e^+e^-\to Z (H_{\mbox{\tiny SM}}(M_{h_1})\to b\bar{b}))} \nonumber \\
& \xi_{\ga}\equiv\frac{\Ga(h_1\to gg)\cdot \br(h_1\to \ga\ga)}{\Ga(H_{\mbox{\tiny SM}}(M_{h_1})\to gg)\cdot \br(H_{\mbox{\tiny SM}}(M_{h_1})\to \ga\ga)}\sim\frac{\sig(gg\to h_1\to \ga\ga)}{\sig(gg\to H_{\mbox{\tiny SM}}(M_{h_1})\to\ga\ga)}\ .
  \label{eq:xi}
\end{align}
These definitions of~$\xi_{b,\ga}$ give estimates of the
signals that~$h_1$ would generate in the~LEP~searches
for~\mbox{$e^+e^-\to Z (H\to b\bar{b})$} and the~LHC~searches
for~\mbox{$pp\to H\to\ga\ga$}, normalized to
the~SM~cross-sections. In the analysis in \citere{Domingo:2018uim}
constraints from ``other sectors'' (such as Dark Matter or $(g-2)_\mu$)
are not taken into account, as they are not closely related to Higgs
sector physics.

The NMSSM parameters are chosen as (see \citere{Domingo:2018uim} for
definitions and details), 
\begin{align}
&\la = 0.6, \quad
\kappa = 0.035, \quad
\tb = 2, \quad
\MHp = 1000 \gev, \quad
A_\kappa = -325 \gev, \nonumber \\
&\mu_{\mbox{\tiny eff}} = (397 + 15 \cdot x) \gev
~(x~\mbox{is varied in the interval}~[0,1]), \nonumber \\
&\mbox{the third generation squark mass scale}~m_{\tilde Q} = 1000 \gev, 
\At = \Ab = 0. \nonumber
\end{align} 
In our analysis we vary~$\mu_{\mbox{\tiny eff}}$ in a narrow interval as
indicated above.
It was tested with 
\texttt{HiggsBounds} \texttt{-4.3.1}
(and~\texttt{5.1.1beta})\,\cite{Bechtle:2008jh,Bechtle:2011sb,Bechtle:2013wla,Bechtle:2015pma,HB-www}
and~\texttt{HiggsSignals-1.3.1}
(and~\texttt{2.1.0beta})\,\cite{Bechtle:2013xfa,Bechtle:2014ewa,HS2,HB-www}
that our parameter points are in agreement with the Higgs rate
measurements at the LHC as well as with the Higgs boson searches at LEP,
the Tevatron and the LHC.

\begin{figure}[htb!]
  \centering
  \includegraphics[width=\textwidth]{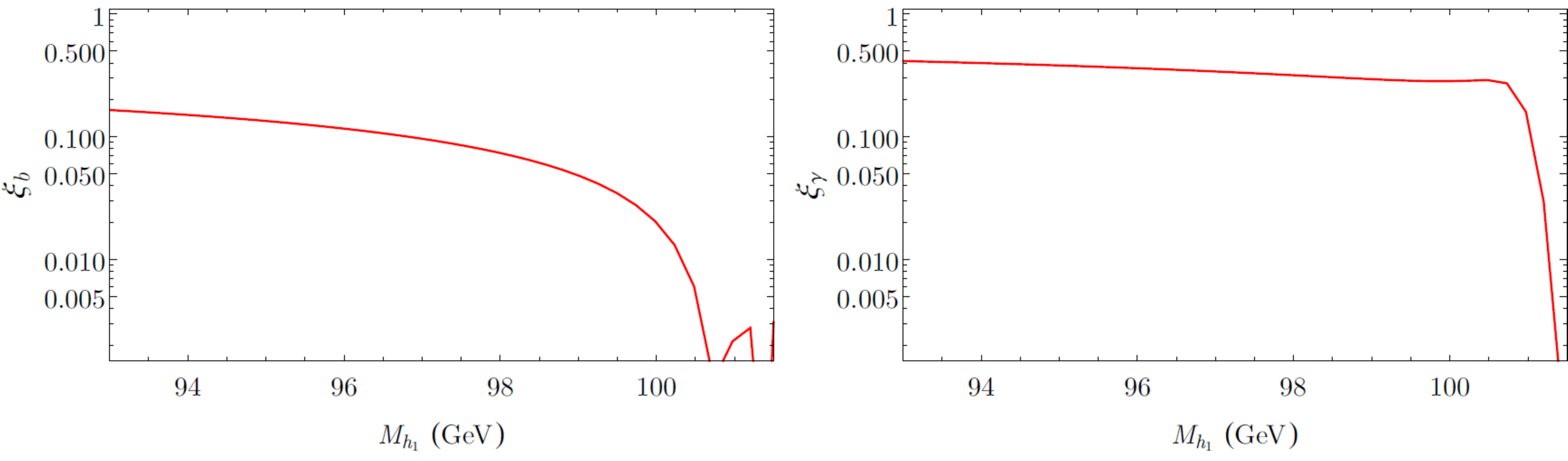}
\vspace{-1em}
  \caption{The
    quantities~$\xi_b$ and~$\xi_{\ga}$ of Eq.\,(3.1),
    estimating the signals associated with~$h_1$ in
    the~$b\bar{b}$~channel at~LEP and in the~$\ga\ga$~channel at
    the~LHC, as compared to an~SM~Higgs at the same mass. Explicit
    values at~\mbox{$M_{h_1}=95.0$}\,GeV are given in
    Tab.\,1.}
\label{fig:NMSSM}
\end{figure}

With growing~$\mu_{\mbox{\tiny eff}}$, the mixing between the two
light~$\cp$-even states increases, eventually pushing the singlet mass
down to~$\sim 90 \gev$ and the mass of the~SM-like state up
to~$\sim 128 \gev$. Consistency with the experimental results
obtained on the observed state at~$125$\,GeV is achieved for a mass of
the~SM-like state that is compatible with the~LHC~discovery within
experimental and theoretical uncertainties. The~$\cp$-odd singlet has a
mass of~$\sim 150 \gev$, while the heavy doublet states are
at~$\sim 1 \tev$ in this scenario. The decay properties of~$h_1$ and
$h_2$ are given in \refta{tab:NMSSM} (for a
specific point). 
The quantities~$\xi_b$ and~$\xi_{\ga}$ of Eq.\,\eqref{eq:xi},
are shown in Fig.~\ref{fig:NMSSM}, 
estimating the signals associated with~$h_1$ in
the~$b\bar{b}$~channel at~LEP and in the~$\ga\ga$~channel at
the~LHC, as compared to an~SM~Higgs at the same mass.
The magnitude of the
estimated~\mbox{$e^+e^-\to Z(h_1\to b\bar{b})$}~signal
reaches~$\sim 13\%$ of that of an~SM~Higgs
at~\mbox{$M_{h_1}\sim 95 \gev$}, 
while~\mbox{$pp\to h_1\to\ga\ga$} corresponds to more than~$40\%$
of an~SM~signal in the same mass range. In this
example,~\mbox{$\text{BR}(h_1\to\ga\ga)$}
(or~\mbox{$\text{BR}(h_1\to gg)$}) is only moderately enhanced with
respect to the~SM~branching fraction due to an~$H_u^0$-dominated
doublet composition of~$h_1$, while~\mbox{$\text{BR}(h_1\to b\bar{b})$} 
remains dominant, albeit slightly suppressed. This
scenario would thus simultaneously address the~LEP and
the~CMS~excesses in a phenomenologically consistent manner.

\begin{table}[htb]
\renewcommand{\arraystretch}{1.2}
\centering
\caption{The Higgs properties for one example point
  in the NMSSM. The Higgs width
  into~$xx'$ is denoted by~$\Ga_{xx'}$, and the width normalized to
  the~SM~width at the same mass is represented
  by~$\widehat{\Ga}_{xx'}$. The symbol~$\widehat{\text{BR}}_{xx'}$
  represents the Higgs branching ratio into~$xx'$, normalized to
  the~SM~branching ratio at the same mass.}  
{\begin{tabular}{|l|cc||l|c|}
\hline
 & $h_1$ & $h_2$ & & $h_1$ \\
\hline
$\Ga_{\ga\ga}$ [GeV] & $6.3\cdot10^{-7}$ & $7.6\cdot10^{-6}$ & 
$\widehat{\text{BR}}_{\ga\ga}$ & $2.2$ \\
$\Ga_{b\bar{b}}$ [GeV] & $1.4\cdot10^{-4}$ & $2.1\cdot10^{-3}$ &
$\widehat{\text{BR}}_{b\bar{b}}$ & $0.88$ \\
$\Ga_{gg}$ [GeV] & $2.2\cdot10^{-5}$ & $2.2\cdot10^{-4}$ &
$\widehat{\Ga}_{gg}$ & $0.18$ \\
$\Ga_{ZZ}$ [GeV] & $1.0\cdot10^{-7}$ & $9.7\cdot10^{-5}$ &
$\widehat{\Ga}_{ZZ}$ & $0.15$ \\
$\Ga_{WW}$ [GeV] & $8.4\cdot10^{-6}$ & $7.9\cdot10^{-4}$ &
$\xi_{\ga}$ & $0.41$ \\
$\Ga_{\tau\tau}$ [GeV] & $1.5\cdot10^{-5}$ & $2.4\cdot10^{-4}$ &
$\xi_b$ & $0.13$ \\
$\Ga_{c\bar{c}}$ [GeV] & $1.8\cdot10^{-5}$ & $1.2\cdot10^{-4}$ & & \\
\hline
$M_{h_i}$ [GeV] & $95.0$ & $125.5$ & & \\
\hline
\end{tabular}}
\label{tab:NMSSM} 
\renewcommand{\arraystretch}{1.0}
\end{table}

\subsection{The \boldmath{\mnSSM} solution}
\label{sec:mnSSM}

The results in this section are based on \citere{Biekotter:2017xmf}. 
Within the \mnSSM\ (in the simplified ``one generation case'')
we will interpret the
light scalar as the $\cp$-even right-handed sneutrino.
Since the singlet of the NMSSM and the right-handed sneutrino of
the \mnSSM\ are both gauge-singlets, they share very similar properties.
However, the explanation of the excesses in the \mnSSM\ avoids
bounds from direct detection experiments, because $R$-parity is broken
in the \mnSSM\ and the dark matter candidate is not a neutralino as in
the NMSSM but a gravitino with a lifetime longer than the age of
the universe~\cite{Munoz:2016vaa}. This can be important since the direct
detection measurements were shown to be very constraining in the NMSSM
while trying to explain the dark matter abundance on top of the
excesses from LEP and CMS~\cite{Cao:2016uwt}.

\begin{table}[b]
\renewcommand{\arraystretch}{1.2}
\centering
\caption{Input parameters for the scenario featuring the right-handed
		 sneutrino in the mass range of the LEP and CMS excesses and
		 a SM-like Higgs boson as next-to-lightest $\cp$-even scalar;
  		 all masses and values for trilinear parameters are in GeV.}  
{\begin{tabular}{|c c c c c c c c c c|}
\hline
 $v_{iL}/\sqrt{2}$ & $Y^\nu_i$ & $A^\nu_i$ & $\tb$ & $\mu$ & $\lambda$ &
 	$A^\lambda$ & $\kappa$ & $A^\kappa$ & $M_1$ \\
 \hline
 $10^{-5}$ & $10^{-7}$ & $-1000$ & $2$ & $[413;418]$ & $0.6$ &
 	$956$ & $0.035$ & $[-300;-318]$ & $100$ \\
 \hline
 \hline
 $M_2$ & $M_3$ & $m_{\widetilde{Q}_{iL}}^2$ &
 	$m_{\widetilde{u}_{iR}}^2$ & $m_{\widetilde{d}_{iR}}^2$ &
 	$A^u_i$ & $A^{d}_i$ & $(m_{\widetilde{e}}^2)_{ii}$ &
 	$A^e_{33}$ & $A^e_{11,22}$ \\
 \hline
 $200$ & $1500$ & $800^2$& $800^2$ & $800^2$ & $0$ &
 	$0$ & $800^2$ & $0$ & $0$ \\
\hline
\end{tabular}}
\label{tab:munuSSM} 
\renewcommand{\arraystretch}{1.0}
\end{table}

In \refta{tab:munuSSM} we list the values of the parameters we used
to account for the lightest $\cp$-even scalar as the right-handed
sneutrino and the second lightest one the SM-like Higgs boson
(see \citere{Biekotter:2017xmf} for definitions and details). 
As in the NMSSM \texttt{HiggsBounds} and \texttt{HiggsSignals} were
used to ensure the compatibility with experimental data.
In the analysis in \citere{Biekotter:2017xmf}
constraints from ``other sectors'' (such as flavor physics or $(g-2)_\mu$)
are not taken into account, as they are not closely related to Higgs
sector physics.
$\lambda$ is chosen to be large to account for a sizable mixing of
the right-handed sneutrino and the doublet Higgs bosons.
In the regime where the SM-like Higgs boson is not the lightest scalar,
one does not need large quantum corrections to the Higgs boson mass
(which were evaluated according to \citere{Biekotter:2017xmf}), 
because the tree-level mass is already well above $100\gev$. This is
why $\tan\beta$ can be low and the soft trilinear couplings $A^{u,d,e}$
are set to zero. The values of $A^{\lambda}$ and $-A^\nu$
are chosen to be around $1\tev$ to get masses for the
heavy MSSM-like Higgs and the left-handed sneutrinos of this order, so
they do not play an important role in the following discussion.
On the other hand, $\kappa$ is small to bring the mass
of the right-handed sneutrino below the SM-like Higgs boson mass.
Finally, the two parameters that are varied are $\mu$ and $A^\kappa$.
By increasing $\mu$ the mixing of the right-handed sneutrino with
the SM-like Higgs boson is increased, which is needed to couple
the gauge-singlet to quarks and gauge-bosons. At the same time we
used the value of $A^\kappa$ to keep the mass of the
right-handed sneutrino in the correct range. 

\begin{figure}[tb!]
  \centering
  \includegraphics[width=0.9\textwidth]{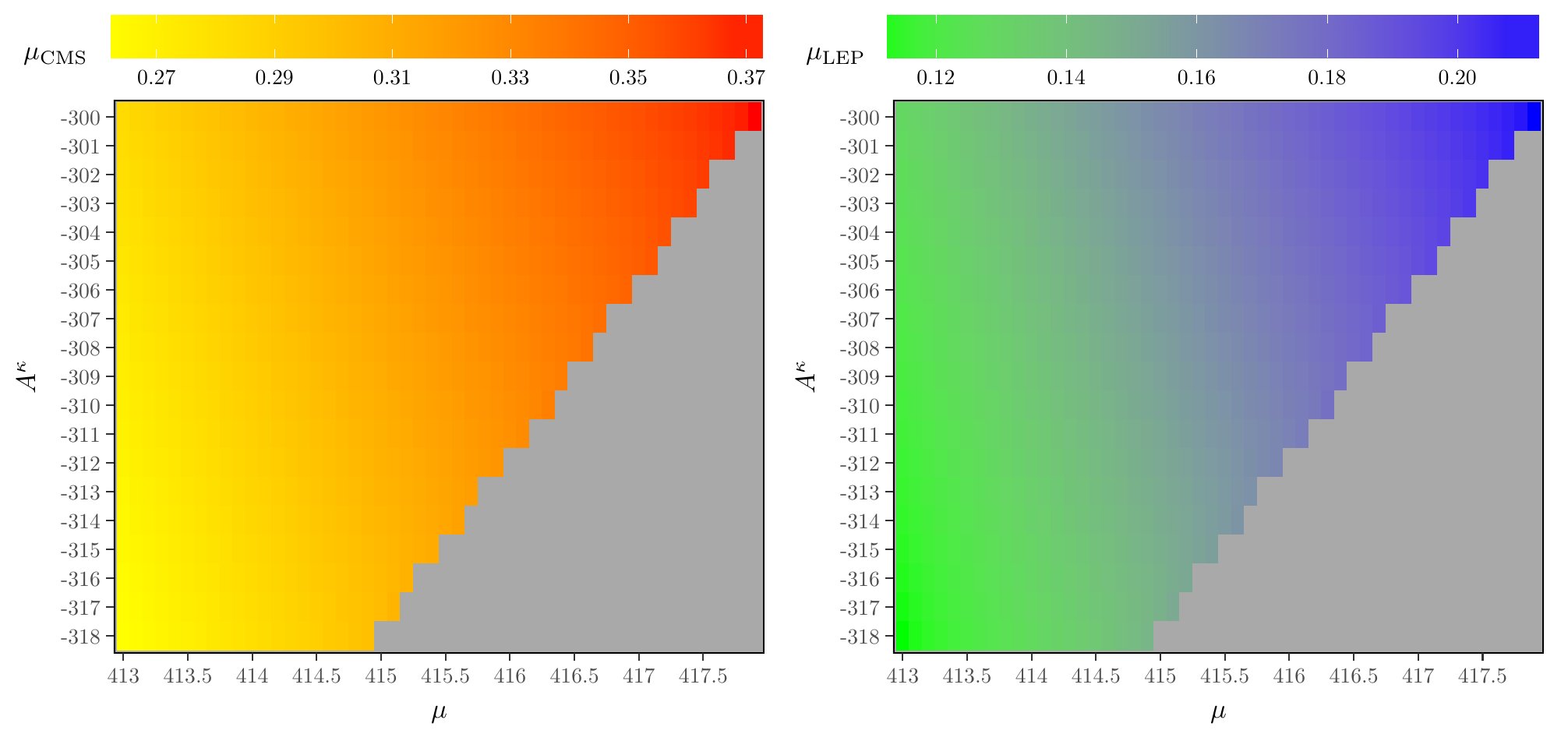}
  \vspace{-1em}
  \caption{Signal strengths for the lightest $\widetilde{\nu}_R$-like
         neutral scalar at CMS ($pp\to h_1 \to \ga\ga$)
         (\textit{left}) and
         LEP (${e^+e^-\to h_1 Z \to b\bar{b} Z}$)
         (\textit{right}) in the
         $\mu$-$A^\kappa$ plane. The gray area is
         excluded because the right-handed sneutrino becomes
         tachyonic at tree-level.}
  \label{fig:munuSSM}
\end{figure}

The result are shown in \reffi{fig:munuSSM}, the CMS (left) and the LEP
excesses (right) in the $\mu$--$A^{\kappa}$ plane.
While the LEP excess is easily reproduced
in the observed parameter space, we cannot achieve the central
value for $\mu_{\rm CMS}$, but only slightly smaller values.
As already observed in \citere{Cao:2016uwt},
the reason for this is that for explaining the LEP excess
a sizable coupling to the bottom quark is needed.
On the contrary, the CMS
excess demands a small value for the $h_1b \bar b$~coupling so that the
total width of the $h_1$ becomes small and $h_1 \to \ga\ga$
is enhanced. Nevertheless, considering the large
experimental uncertainties in $\mu_{\rm CMS}$ and $\mu_{\rm LEP}$,
the scenario reviewed in this section
accommodates both excesses comfortably well
(at approximately $1\sig$).

\section{Conclusions (for future colliders)}
\label{sec:concl}

Searches for Higgs bosons below $125 \gev$ have been performed at LEP,
the Tevatron and the LHC. We have briefly reviewed that 
LEP reported a $2.3\,\sig$ local excess
observed in the~$e^+e^-\to Z(H\to b\bar{b})$
searches\,\cite{Barate:2003sz}, which would be consistent with a
scalar mass of~$\sim 98 \gev$ (but with a rather coarse mass
resolution). Furthermore, recent LHC~Run\,II
results\,\cite{CMS:2017yta} for CMS Higgs searches in the diphoton
final state show a local excess of~$\sim 3\,\sig$ in the vicinity
of~$\sim 96 \gev$, with a similar upward fluctuation
of~$2\,\sig$ in the Run\,I data at a comparable mass.  
First Run\,II~results from~ATLAS
with~$80$\,fb$^{-1}$ in the~$\ga\ga$~searches below~$125$\,GeV are well
compatible with the limit obtained by CMS at $\sim 96 \gev$ (although
not showing a relevant excess).

We have briefly reviewed BSM interpretations, explaining simultaneously
the LEP and the CMS ``excess''. In particular we have reviewed the
solution within the NMSSM and the \mnSSM. Within the NMSSM 
we investigated the 
case of a mostly singlet-like state with a mass 
of~$\lsim 96$\,GeV. The decays of such a state can be notably
affected by suppressed couplings to down- or up-type~quarks which can
occur in certain parameter regions due to the mixing
between the different Higgs states. In particular, an additional Higgs
boson~$h_i$ of this kind could manifest itself via signatures in the
channels~\mbox{$e^+e^- \to Z(h_i\to b\bar{b})$} and/or~\mbox{$pp \to
  h_i \to \ga\ga$}.  The presence of such a light Higgs boson
could thus explain the ``excesses'' reported by~LEP and~CMS in those channels.

Within the \mnSSM\ we reviewed that this model can accommodate a
right-handed ($\cp$-even) scalar neutrino 
with a mass of $\sim 96 \gev$, where the full Higgs sector is in
agreement with the Higgs-boson measurements and exclusion bounds
obtained at the LHC, as well as at LEP and the Tevatron.
It was demonstrated that the light right-handed sneutrino can explain
an excess of $\ga\ga$ events at $\sim 96 \gev$ as reported 
recently by CMS in their Run~I and Run~II date. It can
simultaneously describe the $2\,\sig$ excess of $b \bar b$ events
observed at LEP at a similar mass scale. 

A new Higgs boson with a mass of $\sim 96 \gev$ is easily kinematically
accessible at 
future $e^+e^-$ colliders, assuming an energy of at least 
$\sqrt{s} = 250 \gev$. It was shown in \citere{Drechsel:2018mgd} that a
light Higgs boson explaining the LEP ``excess'' is easily within the
reach of the ILC250. Further confirmation of these ``excesses'' would
strengthen the already robust physics case for such a machine.


\subsection*{Acknowledgements}

\vspace{-0.5em}
S.H.~thanks
T.~Biek\"otter,
F.~Domingo,
C.~Mu\~noz,
S.~Pa\ss ehr
and
G.~Weiglein,
with whom the NMSSM and \mnSSM\ results presented here have been
obtained. S.H.~thanks K.~Tackmann for interesting discussions.
The work of S.H.\ was supported in part by the MEINCOP (Spain) under 
contract FPA2016-78022-P, in part by the Spanish Agencia Estatal de
Investigaci\'on (AEI), in part by
the EU Fondo Europeo de Desarrollo Regional (FEDER) through the project
FPA2016-78645-P, in part by the ``Spanish Red Consolider MultiDark''
FPA2017-90566-REDC, and in part by the AEI through the grant IFT Centro de
Excelencia Severo Ochoa SEV-2016-0597.




\begin{thebibliography}{99}

\vspace{-1em}
\bibitem{Aad:2012tfa}
  G.~Aad {\it et al.} [ATLAS Collaboration],
  Phys.\ Lett.\ B {\bf 716} (2012) 1
  [arXiv:1207.7214 [hep-ex]].

\bibitem{Chatrchyan:2012xdj}
  S.~Chatrchyan {\it et al.} [CMS Collaboration],
  Phys.\ Lett.\ B {\bf 716} (2012) 30
  [arXiv:1207.7235 [hep-ex]].

\bibitem{Khachatryan:2016vau}
  G.~Aad {\it et al.} [ATLAS and CMS Collab.],
  JHEP {\bf 1608} (2016) 045
  [arXiv:1606.02266 [hep-ex]].

\bibitem{Nilles:1983ge}
  H.~P.~Nilles,
  Phys.\ Rept.\  {\bf 110} (1984) 1.

\bibitem{Haber:1984rc}
  H.~E.~Haber and G.~L.~Kane,
  Phys.\ Rept.\  {\bf 117} (1985) 75.

\bibitem{Heinemeyer:2011aa}
  S.~Heinemeyer, O.~St\aa l and G.~Weiglein,
  Phys.\ Lett.\ B {\bf 710} (2012) 201
  [arXiv:1112.3026 [hep-ph]].

\bibitem{Bechtle:2016kui}
  P.~Bechtle, H.~E.~Haber, S.~Heinemeyer, O.~St\aa l, T.~Stefaniak, G.~Weiglein and L.~Zeune,
  Eur.\ Phys.\ J.\ C {\bf 77} (2017) no.2,  67
  [arXiv:1608.00638 [hep-ph]].

\bibitem{Bahl:2018zmf}
  H.~Bahl {\it et al.},
  arXiv:1808.07542 [hep-ph].
  
\bibitem{Ellwanger:2009dp}
  U.~Ellwanger, C.~Hugonie and A.~M.~Teixeira,
  Phys.\ Rept.\  {\bf 496} (2010) 1
  [arXiv:0910.1785 [hep-ph]].

\bibitem{Maniatis:2009re}
  M.~Maniatis,
  Int.\ J.\ Mod.\ Phys.\ A {\bf 25} (2010) 3505
  [arXiv:0906.0777 [hep-ph]].

\bibitem{Ellis:1988er}
  J.~R.~Ellis, J.~F.~Gunion, H.~E.~Haber, L.~Roszkowski and F.~Zwirner,
  Phys.\ Rev.\ D {\bf 39} (1989) 844.

\bibitem{Miller:2003ay}
  D.~J.~Miller, R.~Nevzorov and P.~M.~Zerwas,
  Nucl.\ Phys.\ B {\bf 681} (2004) 3
  [hep-ph/0304049].

\bibitem{King:2012is}
  S.~King, M.~M\"uhlleitner and R.~Nevzorov,
  Nucl.\ Phys.\ B {\bf 860} (2012) 207
  [arXiv:1201.2671 [hep-ph]].

\bibitem{Domingo:2015eea}
  F.~Domingo and G.~Weiglein,
  JHEP {\bf 1604} (2016) 095
  [arXiv:1509.07283 [hep-ph]].

\bibitem{LopezFogliani:2005yw}
  D.~E.~L\'opez-Fogliani and C.~Mu\~noz,
  Phys.\ Rev.\ Lett.\  {\bf 97} (2006) 041801
  [hep-ph/0508297].

\bibitem{Escudero:2008jg}
  N.~Escudero {\it et al.}, 
  JHEP {\bf 0812} (2008) 099
  [arXiv:0810.1507 [hep-ph]].

\bibitem{Munoz:2009an}
  C.~Mu\~noz,
  AIP Conf.\ Proc.\  {\bf 1200} (2010) 413
  [arXiv:0909.5140 [hep-ph]].

\bibitem{Munoz:2016vaa}
  C.~Mu\~noz,
  PoS DSU {\bf 2015} (2016) 034
  [arXiv:1608.07912 [hep-ph]].

\bibitem{Ghosh:2017yeh}
  P.~Ghosh {\it et al.}, 
  Int.\ J.\ Mod.\ Phys.\ A {\bf 33} (2018) no.18n19,  1850110
  [arXiv:1707.02471 [hep-ph]].

\bibitem{Choi:2009ng}
  K.~Y.~Choi, D.~E.~L\'opez-Fogliani, C.~Mu\~noz and R.~Ruiz de Austri,
  JCAP {\bf 1003} (2010) 028
  [arXiv:0906.3681 [hep-ph]].

\bibitem{GomezVargas:2011ph}
  G.~A.~G\'omez-Vargas {\it et al.}, 
  JCAP {\bf 1202} (2012) 001
  [arXiv:1110.3305 [astro-ph.HE]].

\bibitem{Albert:2014hwa}
  A.~Albert {\it et al.} [Fermi-LAT Collab.],
  JCAP {\bf 1410} (2014) no.10,  023
  [arXiv:1406.3430 [astro-ph.HE]].

\bibitem{Gomez-Vargas:2016ocf}
  G.~A.~G\'omez-Vargas {\it et al.}, 
  JCAP {\bf 1703} (2017) no.03,  047
  [arXiv:1608.08640 [hep-ph]].

\bibitem{Barate:2003sz}
  R.~Barate {\it et al.} [ALEPH and DELPHI and L3 and OPAL Collaborations and LEP Working Group for Higgs boson searches],
  Phys.\ Lett.\ B {\bf 565} (2003) 61
  [hep-ex/0306033].

\bibitem{Cao:2016uwt}
  J.~Cao {\it et al.}, 
  Phys.\ Rev.\ D {\bf 95} (2017) no.11,  116001
  [arXiv:1612.08522 [hep-ph]].

\bibitem{Azatov:2012bz}
  A.~Azatov, R.~Contino and J.~Galloway,
  JHEP {\bf 1204} (2012) 127
   Erratum: [JHEP {\bf 1304} (2013) 140]
  [arXiv:1202.3415 [hep-ph]].

\bibitem{CMS:2017yta}
A.~M.~Sirunyan {\it et al.} [CMS Collaboration],
  CMS-PAS-HIG-17-013 
  [arXiv:1811.08459 [hep-ex]].

\bibitem{ATLAS:2018xad}
  The ATLAS collaboration,
  ATLAS-CONF-2018-025.

\bibitem{Heinemeyer:2018jcd}
  S.~Heinemeyer,
  Int.\ J.\ Mod.\ Phys.\ A {\bf 33} (2018) no.31,  1844006.

\bibitem{Fox:2017uwr}
  P.~J.~Fox and N.~Weiner,
  JHEP {\bf 1808} (2018) 025
  [arXiv:1710.07649 [hep-ph]].

\bibitem{Richard:2017kot}
  F.~Richard,
  arXiv:1712.06410 [hep-ex].

\bibitem{Haisch:2017gql}
  U.~Haisch and A.~Malinauskas,
  JHEP {\bf 1803} (2018) 135
  [arXiv:1712.06599 [hep-ph]].

\bibitem{Biekotter:2017xmf}
  T.\,Biek\"otter, S.\,Heinemeyer and C.\,Mu\~noz,
  Eur.\ Phys.\ J.\ C {\bf 78} (2018) no.6, 504
  [arXiv:1712.07475 [hep-ph]].

\bibitem{Liu:2018xsw}
  D.~Liu, J.~Liu, C.~E.~M.~Wagner and X.~P.~Wang,
  JHEP {\bf 1806} (2018) 150
  [arXiv:1805.01476 [hep-ph]].

\bibitem{Domingo:2018uim}
  F.~Domingo, S.~Heinemeyer, S.~Pa\ss ehr and G.~Weiglein,
  arXiv:1807.06322 [hep-ph].

\bibitem{Hollik:2018yek}
  W.~G.~Hollik, S.~Liebler, G.~Moortgat-Pick, S.~Pa\ss ehr and G.~Weiglein,
  arXiv:1809.07371 [hep-ph].

\bibitem{Ellwanger:2010nf}
  U.~Ellwanger,
  Phys.\ Lett.\ B {\bf 698} (2011) 293
  [arXiv:1012.1201 [hep-ph]].

\bibitem{Benbrik:2012rm}
  R.~Benbrik, M.~Gomez Bock, S.~Heinemeyer, O.~St\aa l, G.~Weiglein and L.~Zeune,
  Eur.\ Phys.\ J.\ C {\bf 72} (2012) 2171
  [arXiv:1207.1096 [hep-ph]].

\bibitem{Bechtle:2008jh}
  P.~Bechtle, O.~Brein, S.~Heinemeyer, G.~Weiglein and K.~E.~Williams,
  Comput.\ Phys.\ Commun.\  {\bf 181} (2010) 138
  [arXiv:0811.4169 [hep-ph]].

\bibitem{Bechtle:2011sb}
  P.~Bechtle, O.~Brein, S.~Heinemeyer, G.~Weiglein and K.~E.~Williams,
  Comput.\ Phys.\ Commun.\  {\bf 182} (2011) 2605
  [arXiv:1102.1898 [hep-ph]].

\bibitem{Bechtle:2013wla}
  P.~Bechtle, O.~Brein, S.~Heinemeyer, O.~St\aa l, T.~Stefaniak, G.~Weiglein and K.~E.~Williams,
  Eur.\ Phys.\ J.\ C {\bf 74} (2014) no.3,  2693
  [arXiv:1311.0055 [hep-ph]].

\bibitem{Bechtle:2015pma}
  P.~Bechtle, S.~Heinemeyer, O.~St\aa l, T.~Stefaniak and G.~Weiglein,
  Eur.\ Phys.\ J.\ C {\bf 75} (2015) no.9,  421
  [arXiv:1507.06706 [hep-ph]].

\bibitem{HB-www} {\tt http://higgsbounds.hepforge.org}~.

\bibitem{Bechtle:2013xfa}
  P.~Bechtle, S.~Heinemeyer, O.~St\aa l, T.~Stefaniak and G.~Weiglein,
  Eur.\ Phys.\ J.\ C {\bf 74} (2014) no.2,  2711
  [arXiv:1305.1933 [hep-ph]].

\bibitem{Bechtle:2014ewa}
  P.~Bechtle, S.~Heinemeyer, O.~St\aa l, T.~Stefaniak and G.~Weiglein,
  JHEP {\bf 1411} (2014) 039
  [arXiv:1403.1582 [hep-ph]].

\bibitem{HS2} P.~Bechtle, D.~Derks, S.~Heinemeyer, T.~Klingl, T.~Stefaniak,
  G.~Weiglein, IFT-UAM/CSIC-18-125.

\bibitem{Drechsel:2018mgd}
  P.~Drechsel, G.~Moortgat-Pick and G.~Weiglein,
  arXiv:1801.09662 [hep-ph].
  
\end{thebibliography}
\end{document}